\documentclass[twocolumn,amsmath,amssymb,nofootinbib,floatfix,prl,superscriptaddress]{revtex4}

\usepackage{graphicx}
\usepackage{bm, bbm}      

\begin{document}

\title{Fractional statistics and duality: strong tunneling behavior
of edge states of quantum Hall liquids in the Jain sequence}

\author{Claudio Chamon}
\affiliation{
Physics Department, Boston University, 
590 Commonwealth Ave., Boston, MA 02215, USA}

\author{Eduardo Fradkin}
\affiliation{
Department of Physics, University of Illinois at Urbana-Champaign, 
1110 W. Green St., Urbana IL 61801-3080, USA}

\author{Ana L\'opez}
\affiliation{
OUCE, Oxford University, 
South Parks Rd, Oxford OX1 3QY, UK
}

\date{\today} 

\begin{abstract}
  While the values for the fractional charge and fractional statistics
  coincide for fractional Hall (FQH) states in the Laughlin sequence,
  they do not for more general FQH states, such as those in the Jain
  sequence. This mismatch leads to additional phase factors in the
  weak coupling expansion for tunneling between edge states which
  alter the nature of the strong tunneling limit. We show here how to
  construct a weak-strong coupling duality for generalized FQH states
  with simple unreconstructed edges. The correct dualization of
  quasiparticles into integer charged fermions is a consistency
  requirement for a theory of FQH edge states with a simple edge.  We
  show that this duality also applies for weakly reconstructed edges.
\end{abstract}
\maketitle

The existence of excitations with fractional statistics, both Abelian
and non-Abelian, is one of the most startling and unavoidable
predictions of the theory of the fractional quantum (FQH) Hall
effect~\cite{Halperin,Arovas,Moore-Read1991}. Information on the quasiparticle
statistics can in principle be extracted using tunneling and
interference experiments. Consequently, a number of proposals for the
measurement of fractional statistics of FQH fluids have been made in
recent years. These schemes consist of tunneling interferometers which
make use of the properties of the edge state tunneling in the FQH
effect~\cite{CFKSW,Safi,Kane,Smitha,Kim,Kim06}. Two recent
experiments, using an interferometer in which $\nu=1/3$ quasiparticles
can enclose an island with filling fraction $\nu=2/5$, reported
observations consistent with fractional quasiparticle
statistics and with chiral Luttinger liquid dependence
of the thermal dephasing of the Aharonov-Bohm
oscillations~\cite{Goldman1}.  

In most theoretical proposals for detecting fractional statistics,
calculations are carried out at the lowest orders in perturbation
theory in the tunneling amplitude at a point contact. For the Laughlin
states the problem of a single point contact can be fully solved under
the assumption that the edge remains sharp (i.e. it is
unreconstructed). As the tunneling amplitude increases, the dynamics
of the point contact evolves from being dominated by quasiparticle
tunneling at weak coupling to a strong coupling regime in which the
fluid is pinched off and electron tunneling
dominates\cite{Wen1991,Kane-Fisher2}. This process is embodied by a
powerful electron-quasiparticle
duality\cite{Kane-Fisher2,Fendley-etal1} which also encodes the
information on the quantum numbers of the excitations. Since for the
Laughlin states the filling factor, the fractional charge and the
fractional statistics are essentially the same number, it is not
possible to glean direct evidence for fractional quantum numbers of
the excitations directly from the $I-V$ curves at a single point
contact beyond the observation that the FQH edges are chiral Luttinger
liquids (for a review on FQH tunneling see Ref. \cite{Chang2003} and
references therein.)

For generic FQH states the situation is more complex and it is not
clear if a unique duality exists (Recently\cite{Fendley2006}, a dual
picture was proposed for the non-Abelian FQH state at $\nu=5/2$.)
Within the hierarchical description of the FQH edge states (for a
review see Ref.~\cite{Wen1995}), the electron droplet acquires an
onion-shell-like structure, with the states lower in the hierarchy
laying in the outer regions and the higher states in the inner
regions. Given this structure, a point contact can (and does) mix
these edges in a complex way. Thus one can envision a series of
pinch-off transitions as one layer is peeled-off after
another. Clearly, there is no universal electron-quasiparticle duality
in this regime.

On the other hand, if the edge of a hierarchical FQH state were to
remain sharp down to the scale of the magnetic length, it should not
be possible (or meaningful) to physically resolve the many layers
assumed by the hierarchies. In this regime the edge should become
simple. Such a theory of a simple unreconstructed edge state for the
Jain FQH states was presented in Ref.~\cite{LF}. The main purpose of
this paper is to show that for these simple edges there is an uniquely
defined electron-quasiparticle duality, representing the pinch-off
process. Furthermore, it will turn out that this construction is
compatible with a degree of edge reconstruction.

A straightforward extension of duality to the case of the Jain states
is hindered by differences in the structure of the Coulomb gas
expansion (discussed below) in the form of extra statistical phase
factors for the tunneling events around each of the starting fixed
points (quasiparticle and electron tunneling). These phase factors
spoil the simple weak tunneling {\it vs.}  instanton duality clearly
at work in the Laughlin states.  Here we show how to use the structure
of these expansions to construct a generalized duality for the edge
states of the Jain sequence.  A key ingredient in the construction of
this generalized duality is to account for the correct statistics of
the quasiparticles, which is indeed {\it necessary} in order to reach
the electron tunneling limit (with the correct dual fermionic
statistics).  To achieve this goal, we combine the edge state
formulation for the Jain states in Ref.~\cite{LF} to recent advances
on treating statistical phases in quantum impurity problems in
Ref.~\cite{COA1,OCA2}.

The formulation of edge modes of Ref.~\cite{LF} contains a charged
mode, and a neutral topological mode. The real-time Lagrangian is
\begin{eqnarray}
{\cal S}_{R/L}& =&  {\frac {1}{4\pi}} 
\int dt\,dx\;
\partial_x\phi^C_{R,L}\;
\left(\mp \partial_t-\partial_x\right)\phi^C_{R,L} \nonumber\\
& +&  {\frac {1}{4\pi}} 
\int dt\,dx\;
\partial_x\phi^T_{R,L}\;
\left(\pm \partial_t\right)\phi^T_{R,L}
\;,
\label{eq:S-charge}
\end{eqnarray}
where the velocity of the charge mode is set to unit. In real time, the
propagators for these charged modes are:
\begin{eqnarray}
\langle\phi_{R,L}^C(x,t)\; \phi_{R,L}^C (0,0)\rangle
&=&- \ln[\delta+i(t\mp x)]  \nonumber\\
&& \!\!\!\!\!\!\!\!\!\!\!\!\!\!\!\!\!\!\!\!\!\!\!\!
\to - \ln |t\mp x| - i {\frac \pi 2}\; {\rm sgn}(t\mp x) 
\,,
\label{eq:propC}
\end{eqnarray}
where we considered $|t|,|x|\gg\delta$, a UV cut-off. The real-time propagators for the
topological neutral modes are
\begin{eqnarray}
\langle\phi_{R,L}^T(x,t) \;\phi_{R,L}^T (0,0)\rangle = i {\frac \pi
2} \; {\rm sgn}(t+\varepsilon)\;{\rm sgn}(\varepsilon \mp x) \;.
\label{eq:propT}
\end{eqnarray}
where $\varepsilon\to0_+$ is a regulator such that for $x=0$ or $t=0$
results $\rm sgn(\epsilon)=1$, ensuring that the propagator is
consistent with the equal-time commutation relations of the
topological fields, and the statistics of the quasiparticles~\cite{note}.

Edge quasiparticle and electron operators can be constructed as vertex
operators:
\begin{equation}
\Psi_{R/L}(x,t) \sim 
e^{\pm i\,\gamma_C\; \phi_{R,L}^C(x,t)} \;\;e^{\pm i\,\gamma_T \;\phi_{R,L}^T(x,t)}
\,.
\label{eq:vertex}
\end{equation}
The charge and statistics of the particles are given by
$e^*=e\sqrt{\nu}\,\gamma_C$ and $\theta/\pi=\gamma_C^2-\gamma_T^2$. Hence, 
for quasiparticles with charge $\nu=p/(2np+1)$, $\gamma^{\rm
qp}_C=\sqrt{\nu}/p$ and $\gamma^{\rm qp}_T=\sqrt{1+1/p}$, and for
electron operators $\gamma^{\rm e}_C=1/\sqrt{\nu}$ and $\gamma^{\rm
e}_T=\sqrt{p^2+p/\nu^2}$~\cite{LF}.

Tunneling operators that move one quasiparticle or electron between
$L$ and $R$ edges at a tunneling point at $x=0$ can be written as
\begin{equation}
T^\pm(t)=\Psi_{L,R}^\dagger \Psi_{R,L}\big|_{x=0}
\sim 
e^{\pm i\,\gamma_C\; \phi^C(t)} \;\;e^{\pm i\,\gamma_T \;\phi^T(t)}
\,,
\label{eq:T}
\end{equation}
with $\phi^{C,T}(t)=\phi_{R}^{C,T}(0,t)+\phi_{L}^{C,T}(0,t)$. It is
convenient at this point to switch to imaginary-time, in which case
the propagators at $x \rightarrow 0$ become
\begin{eqnarray}
\langle\phi^C(\tau)\; \phi^C (0)\rangle
&=& - \ln |\tau|^2\\
\langle\phi^T(\tau) \;\phi^T (0)\rangle
&=& + i {\pi} \; {\rm sgn}(\tau)
\label{eq:prop-imag-time}
\end{eqnarray}

Tunneling process enter in the action through a term of the form
\begin{equation}
{\cal S}_{\rm tun} =  \int d\tau\;
\left[\Gamma\;T^+(\tau)+\Gamma^*\;T^-(\tau)\right]
\;.
\label{eq:S-tun}
\end{equation}
where $\Gamma$ is the tunneling amplitude and $\Gamma^*$ its complex
conjugate.  A perturbative series in the tunneling amplitude for
quasiparticles corresponds to a Coulomb gas expansion using insertions
of $T^\pm(\tau_i)$ at times labeled by $\tau_i$, containing, at each
order in the expansion of the partition function, terms such as
\begin{eqnarray}
\lefteqn{\langle T^{q_n}(\tau_n) \; \ldots \;T^{q_2}(\tau_2) \;
        T^{q_1}(\tau_1) \rangle
= } \nonumber \\
&\delta(\sum_j q_j) &  \exp{ \left[  \gamma_C^2 
    \sum_{j>k} q_j \,q_k
      \ln{|\tau_j - \tau_k|}^2 \right. }   \nonumber \\
&&  \!\!\!\!\!\!\!\!
\left.
-i \pi \;\gamma_T^2 \sum_{j>k} q_j \,q_k
\;\mathrm{sgn}(\tau_j - \tau_k) \right]
\;,
\label{eq:Coulombgas1}
\end{eqnarray}
where $q_i=\pm 1$ is the charge associated with each vertex operator
that is inserted. The $\sum_j q_j=0$ is the neutrality condition of
the Coulomb gas expansion (non-neutral terms give vanishing
expectation values).

An important feature of the expansion
presented in Eq.\eqref{eq:Coulombgas1} is that, even in the
imaginary-time formulation, there are pure phase factors coming from
the contributions due to the topological modes. In the Laughlin
sequence, only the real terms (those with the logs) are present, and
the weak-strong tunneling duality simply corresponds to an
electric/magnetic charge duality~\cite{Kane-Fisher2}. In the Jain
sequence, the extra phase factors change order by order the prefactors
in the perturbative expansion of the partition function, and
consequently change the nature of the strong fixed point reached at
large tunneling amplitude $\Gamma$. The situation is similar
to that found in Ref.~\cite{COA1,OCA2}, where phase factors in the
Coulomb gas expansion due to fermionic statistics alter qualitatively
the strong coupling limit.

In order to take into account the phase factors due to the topological
modes in Eq.~(\ref{eq:Coulombgas1}), we construct a different
representation of the problem which gives, order by order, the same
factors in the Coulomb gas expansion. In this alternative
representation the weak-strong duality becomes apparent.  The same
phase factors in Eq.~(\ref{eq:Coulombgas1}) can be accounted for in
the following quantum impurity problem, which involves a bosonic
$\varphi$-field and its conjugate momentum, the dual
$\theta$-field. Both fields enter when writing the vertex operators
that represent the tunneling operators.  Consider the tunneling
operator
\begin{equation}
{\tilde T}^q(\tau,x)\sim
e^{iq\,\alpha \varphi(\tau,x)} \;\;
e^{iq\,\beta \theta(\tau,x)},
\end{equation}
where $q=\pm 1$ is the charge associated with this composite vertex
operator. The correlation function of $n$ such operators in the
Coulomb gas expansion is
\begin{equation} 
\langle {\tilde T}^{q_n}(\tau_n,x_n) \;\ldots\;
        {\tilde T}^{q_2}(\tau_2,x_2)\;
        {\tilde T}^{q_1}(\tau_1,x_1) \rangle .
\end{equation}
In evaluating the expression above, we will take the coordinate $x$ to
play the role of time (we will use $x$-ordering, $x_1<x_2<\dots<x_n$,
and all of these $\to 0$), and take $\tau$ to be in the space
direction, so that the canonical commutation relation reads
\begin{equation}
[\theta (\tau) , \varphi(\tau')] =
i \pi \;{\mathrm{sgn}}(\tau - \tau').
\end{equation}

Now, move all the exponentials of ${\theta}$ to the right, using the
commutation relations. Each time one moves an exponential of
${\theta}$ past an exponential of $\varphi$ there is an extra phase
factor. The total phase accumulated is
\begin{equation}
e^{ - i \pi\;\alpha\beta
\sum_{j>k} q_j \,q_k \; {\mathrm{sgn}}(\tau_j - \tau_k)}.
\label{eq:Klein}
\end{equation}
Once moved to the right, the exponentials of $\theta$ act like the
identity operator, {\it i.e.\/} they give a factor of $1$, when
applied to a state $|D\rangle$ with Dirichlet boundary condition (BC)
on $\theta$. This is so since $\theta |D\rangle= \theta_0 |D\rangle$,
where $\theta_0$ is a constant, and the neutrality of the Coulomb gas
ensures that exponentials of all the $\theta$'s at different times
multiply to 1.

The remaining correlation functions of the exponentials of $\varphi$
can be calculated with respect to the state $|D\rangle$, and together
with the extra phase factors, give:
\begin{eqnarray}
\lefteqn{\langle {\tilde T}^{q_1}(\tau_1,x_1) \;{\tilde T}^{q_2}(\tau_2,x_2) \; \ldots\;
        {\tilde T}^{q_n}(\tau_n,x_n) \rangle_D
= } \nonumber \\
&\delta(\sum_j q_j) &  \exp{ \left[  \alpha^2
    \sum_{j>k} q_j \,q_k
      \ln{|\tau_j - \tau_k|}^2 \right. }   \nonumber \\
&&  \!\!\!\!\!\!\!\!
\left.
-i \pi \;\alpha\beta \sum_{j>k} q_j \,q_k
\;\mathrm{sgn}(\tau_j - \tau_k) \right].
\label{D}
\end{eqnarray}

A similar calculation can be carried for the case in which $\theta$
obeys a Neumann boundary condition (or Dirichlet on $\varphi$, {\it
i.e.\/}, $\varphi |N\rangle= \varphi_0 |N\rangle$). In this case, the
exponentials of $\varphi$ that were moved to the left give a factor of
1 when applied to the $\langle N|$ state. Such calculation gives:
\begin{eqnarray}
\lefteqn{\langle {\tilde T}^{q_1}(\tau_1,x_1) \;{\tilde T}^{q_2}(\tau_2,x_2)\;  
\ldots\;        {\tilde T}^{q_n}(\tau_n,x_n) \rangle_N
= } \nonumber \\
&\delta(\sum_j q_j) &  \exp{ \left[  \beta^2
    \sum_{j>k} q_j \,q_k
      \ln{|\tau_j - \tau_k|}^2 \right. }   \nonumber \\
&&  \!\!\!\!\!\!\!\!
\left.
-i \pi \;\alpha\beta \sum_{j>k} q_j \,q_k
\;\mathrm{sgn}(\tau_j - \tau_k) \right].
\label{N}
\end{eqnarray}
We can now compare the expansions for the tunneling problem of Jain
quasiparticles and electrons in Eq.~(\ref{eq:Coulombgas1}) to those in
Eqs.~(\ref{D},\ref{N}). We obtain an identity if we take
\begin{equation}
\label{eq:parameters}
\alpha=\frac{\sqrt\nu}{p}
\qquad
\beta=\frac{1}{\sqrt\nu}
\;,
\end{equation}
for which we obtain a quasiparticle-electron duality.

{\em Dirichlet BC: the quasiparticle}. With the state $|D\rangle$, the
scaling dimension (the prefactor of the log) is $({\gamma^{\rm
    qp}_C})^2=\alpha^2=\nu/p^2$. The extra phase factor is $-\pi
\alpha\beta=-\pi \frac{1}{p}$. These factors must be compared to the phase
factors in Eq.~(\ref{eq:Coulombgas1}), which are $-\pi ({\gamma^{\rm
    qp}_T})^2= -\pi \left(\frac{1}{p}+1\right)$. The extra $-\pi \times 1$
can be simply obtained by using an extra Majorana fermion in front
of the $\tilde T$ operator. Thus, we match the series expansion in
Eq.~\ref{D} for the case of $D$ BC on the field $\varphi$ with the series
expansion for the quasiparticle tunneling problem.

{\em Neumann BC: the electron}.  With the state $|N\rangle$, the
scaling dimension (the prefactor of the log) is $({\gamma^{\rm
e}_C})^2=\beta^2=1/\nu$. The extra phase factor is $-\pi
\alpha\beta=-\pi \frac{1}{p}$. One thus matches the result from the
expansion in the case of electron tunneling, in which case the phase
factors are $-\pi ({\gamma^{\rm e}_T})^2=-\pi
\frac{p^2+p}{\nu^2}\equiv -\pi \left(\frac{1}{p}+1\right)\;({\rm
mod}\,2\pi)$. Again, we are just missing an extra $\pi \times 1$ that
can be simply obtained by using an extra Majorana fermion in front of
the $T$ operator.

What we learn from the mapping between the tunneling problem for the
Jain states and the auxiliary quantum impurity problem is that the
extra phase factor from the topological modes introduced in Ref.~\cite{LF}
can be accounted for by keeping both the field $\varphi$ and its
conjugate momentum $\theta$ in the vertex operator. Once
one chooses the correct vertex operators with both $\varphi$ and
$\theta$, duality amounts to swapping the boundary conditions $N$
and $D$.

Thus, we have constructed an explicit strong tunneling-weak tunneling
duality transformation for generic states in the Jain sequence. The
transformation exchanges the quasiparticle and the electron,
and Dirichlet with Neumann boundary conditions, and can be summarized
by the mapping $\frac{\sqrt{\nu}}{p} \leftrightarrow
\frac{1}{\sqrt{\nu}}$. It is consistent with the well known duality
for the Laughlin states ($p=1$).

All of the discussion above concerns a sharp edge, containing
simply a charge mode and a topological mode. A weakly reconstructed
edge will contain additional neutral modes. (We will not discuss here the generic reconstructed case which is highly non-universal.) These neutral
modes, and the local details of the contact, alter the scaling dimentions of the tunneling
operators. These changes must be balanced by additional phases in
order to provide the correct statistics at a given starting fixed
point. Indeed, two interesting question are the following: 1) How to
dualize a theory once neutral modes are added, given that the correct
statistics are fixed in the starting point? 2) Does the dualized
theory contains quasiparticles with the correct statistics at the
other endpoint? We answer both questions below.

Generically, for tunneling through a single
point (single impurity problem), the neutral modes and the charge mode
can be consolidated together into a single field $\phi^P(t)$.  This
field encapsulates (for the purpose of tunneling) all the {\it propagating}
modes, including the effects of interactions between them. The new
tunneling operator can be obtained by replacing $\phi^C\to\phi^P$ with
$\gamma^2_C\to\gamma^2_P=(1+\epsilon)\;\gamma^2_C$ into
Eq.~(\ref{eq:T}) for $T^\pm(t)$. Notice that the scaling dimension of
the tunneling operator is changed to $(1+\epsilon)\;\gamma^2_C$.

The statistical angle of the tunneling quasiparticles is:
\begin{equation}
{\theta}/{\pi}=\gamma^2_C\;(1+\epsilon)-\gamma^2_T,
\label{eq:angle}
\end{equation}
which requires picking a coefficient for the topological terms such
that $\epsilon\,\gamma^2_C-\gamma^2_T=-\bar\gamma^2_T$, with
$\bar\gamma_T$ the original coefficient in the absence of the neutral
modes. Such choice will now yield, in imaginary time, terms such as
\begin{eqnarray}
\lefteqn{\langle T^{q_n}(\tau_n) \; \ldots \;T^{q_2}(\tau_2) \;
        T^{q_1}(\tau_1) \rangle
= } \nonumber \\
&\delta(\sum_j q_j) &  \exp{ \left[  \gamma_C^2\;(1+\epsilon)
    \sum_{j>k} q_j \,q_k
      \ln{|\tau_j - \tau_k|}^2 \right. }   \nonumber \\
&&  \!\!\!\!\!\!\!\!
-i \pi \;\epsilon\,\gamma_C^2 \sum_{j>k} q_j \,q_k
\;\mathrm{sgn}(\tau_j - \tau_k) 
\;\nonumber \\
&&  \!\!\!\!\!\!\!\!
\left.
-i \pi \;\bar\gamma_T^2 \sum_{j>k} q_j \,q_k
\;\mathrm{sgn}(\tau_j - \tau_k) \right]
\;
\label{eq:Coulombgas2}
\end{eqnarray}
in the Coulomb gas expansion of tunneling events. The third line in
Eq.~(\ref{eq:Coulombgas2}) is the phase that we can represent in terms
of Eq.~(\ref{eq:Klein}), which originated from moving the $\varphi$
fields leftward and the $\theta$ fields rightward, past one another,
in the reordering of the vertex operators. It is the first and second
lines in Eq.~(\ref{eq:Coulombgas2}) that has to be dealt with
now. These two terms together can be accounted for if the equivalent
dissipative quantum mechanics action ${\cal S}_{\rm diss}$ for the
$\phi$ field is changed as follows:
\begin{equation}
\frac {1}{4\pi}\int\!\!\! d\omega |\omega|\;|\varphi(\omega)|^2
\to
\frac {1}{4\pi}\int \!\!\!d\omega\; (A |\omega| + B\omega)\;|\varphi(\omega)|^2
\;.
\label{eq:dissipative-phi}
\end{equation}
With such dissipative action, one obtains the following correlation
functions for the $\varphi$ and $\theta$ fields:
\begin{eqnarray}
&&\!\!\!\!\!\!\!\!\!\!\!\!\!\!\!\langle\varphi(\tau)\, \varphi (0)\rangle
= -\frac{A}{A^2-B^2}\ln |\tau|^2 
+ i{\pi}\frac{B}{A^2-B^2}  {\rm sgn}(\tau)
\label{eq:prop-diss-phi}
\\
&&\!\!\!\!\!\!\!\!\!\!\!\!\!\!\!\langle\theta(\tau)\, \theta (0)\rangle
=-{A}\ln |\tau|^2 
- i{\pi}{B}\;  {\rm sgn}(\tau)
\label{eq:prop-diss-theta}
\;.
\end{eqnarray}
Starting from quasiparticle tunneling, one can match the Coulomb gas
expansion in Eq.~(\ref{eq:Coulombgas2}) by choosing the same
$\alpha,\beta$ as before [which ensures that the third line in
Eq.~(\ref{eq:Coulombgas2}) remains the same], and
\begin{eqnarray}
\frac{A}{A^2-B^2}
=1+\epsilon^{\rm qp}
\;,
\quad
\frac{B}{A^2-B^2}  
=\epsilon^{\rm qp}
\;.
\end{eqnarray}
Notice that from these relations it follows that $A+B=1$, {\it
regardless} of the contribution from the neutral modes. This is {\it
very important}, since the statistics of the particle that tunnels in
the dual picture (the electron) can be obtained using the $\theta$
correlators to be
\begin{equation}
\frac{\theta^e}{\pi}=\beta^2\;(A+B)-\alpha\beta-1
\;
\label{eq:angle-electron}
\end{equation}
(the $-1$ is due to the Majorana modes) which is then {\it
independent} of the neutral modes added to the dual quasiparticle
tunneling fixed point. For completeness, let us list the relations
between the $A,B$ coefficients and the electron parameters:
\begin{eqnarray}
{A}
=
1+\epsilon^{\rm e}
\;,
\quad
-{B}
=
\epsilon^{\rm e}
\;,
\end{eqnarray}
{}from which it follows a relation $(\epsilon^{\rm
e})^{-1}+(\epsilon^{\rm qp})^{-1}=-2$. In turn, this constraints the
relation between the exponents of the $I-V$ tunneling
characteristics at the two regimes:
\begin{equation}
I \propto V^{2g-1}
\label{eq:I-V}
\end{equation}
where $g=\gamma^2_C (1+\epsilon)$. The
exponents associated with quasiparticle and electron tunneling satisfy
\begin{equation}
\frac{1/\nu}{g^{\rm e}}+\frac{\nu/p^2}{g^{\rm qp}}=2
\;.
\label{eq:exponent-relation}
\end{equation}
[Notice that in the absence of neutral modes, $g^{\rm e}=\nu^{-1}$ and
$g^{\rm qp}=\nu/p^2$, and Eq.~(\ref{eq:exponent-relation}) is trivially
satisfied.]

We presented a construction of a quasiparticle-electron duality valid
for generic states of the Jain sequences of FQH states with sharp unreconstructed edges. we showed that in this case too there is a strong tunneling-weak tunneling
duality. A key ingredient of this nonperturbative construction is the
crucial role played by fractional statistics as a consistency
condition.  The weak-to-strong tunneling crossover strongly reflects
the fractional statistics of the FQH quasiparticles. 
This mapping also applies for the case of a weakly reconstructed edge.

\noindent
{\bf Acknowledgments}: This work was supported in part by the National
Science Foundation through the grants DMR 0305482 (CC) at Boston
University, and DMR 0442537 (EF) at the University of Illinois.

\bibliography{composite.bib}

\begin{thebibliography}{20}
\bibitem[{not()}]{note}
\bibinfo{note}{The Lagrangian for the topological fields, the second term of
  Eq. \eqref{eq:S-charge}, dictates their equal time commutation relations. It
  has a huge symmetry as it is invariant under the shift of the fields by
  arbitrary functions of time and space separately. As such, it is not
  invertible, and the propagator of Eq. \eqref{eq:propT} is defined by a set of
  boundary conditions fixed by the regulator $\epsilon$, which ensures the
  consistency with the commutation relations.}
\end{thebibliography}

\end{document}